\documentclass[onecolumn,prd,aps,amsmath,showpacs,amssymb]{revtex4}

\usepackage[latin1]{inputenc}
\usepackage{graphicx}
\usepackage{subfigure}
\usepackage{color}
\usepackage{dcolumn}
\usepackage{bm}
\usepackage{latexsym}
\usepackage{amsmath}
\usepackage{amsfonts}
\usepackage{amssymb}
\newcommand{\VEV}[1]{\langle \overline{#1} {#1} \rangle}
\newcommand{\vev}[1]{\langle #1 \rangle}

\newcommand{\TeV}{\, \text{TeV}}
\newcommand{\GeV}{\, \text{GeV}}

\newcommand{\crit}{{\rm crit}}

\newcommand{\gtwo}{I\kern-.1em I\,}
\newcommand{\beq}{\begin{eqnarray}}
\newcommand{\eeq}{\end{eqnarray}}
\newcommand{\bpm}{\begin{pmatrix}}
\newcommand{\epm}{\end{pmatrix}}
\newcommand{\cl}{\, \rm C.L.}

\newcommand{\drawsquare}[2]{\hbox{%
\rule{#2pt}{#1pt}\hskip-#2pt
\rule{#1pt}{#2pt}\hskip-#1pt
\rule[#1pt]{#1pt}{#2pt}}\rule[#1pt]{#2pt}{#2pt}\hskip-#2pt
\rule{#2pt}{#1pt}}

\newcommand{\Ysymm}{\raisebox{-.5pt}{\drawsquare{6.5}{0.4}}\hskip-0.4pt%
        \raisebox{-.5pt}{\drawsquare{6.5}{0.4}}}

\begin{document}

\title{Topcolor-like dynamics and new matter generations}

\author{Hidenori S. Fukano}
\email{hidenori.f.sakuma@jyu.fi} 
\author{Kimmo Tuominen}
\email{kimmo.i.tuominen@jyu.fi}
\affiliation{Department of Physics, University of Jyv\"askyl\"a, P.O.Box 35, FIN-40014 Jyv\"askyl\"a, Finland \\
and 
Helsinki Institute of Physics, P.O.Box 64, FIN-00014 University of Helsinki, Finland\\}

\begin{abstract}
We explore a scenarios where topcolor-like dynamics operates in the presence of fourth generation matter fields. Using the Minimal Walking Technicolor as a concrete basis for model building, we construct explicit models and confront them with phenomenology. We show that if a new QCD generations exist, both the top-bottom mass splitting as well as the splitting between bottom quark mass and the masses of the fourth generation quarks can be naturally explained within topcolor-like dynamics. On the other hand, the much studied Minimal Walking Technicolor model where only a fourth generation of leptons arise, also leads to a viable model.
\end{abstract}

\maketitle

\section{Introduction}

Even if the Standard Model (SM) of particle interactions passes a large number of experimental tests, we know that it cannot be the ultimate model of nature. This is so, since SM fails to explain the origin of the observed mass patterns of the matter fields, the origin of matter-antimatter asymmetry and the abundance of cold dark matter.

Several extensions of the SM have been proposed. Recently, there has been a lot of developments in model building utilizing strong dynamics. Generally, in Technicolor models the Higgs sector of the SM is declared to be a low energy effective theory in which the Higgs is not elementary but composite. Electroweak symmetry breaking is triggered by a dynamical low energy condensate of new matter fields called technifermions. The mechanism is automatically insensitive to high energy physics; see \cite{Sannino:2009za,Sannino:2008ha,Hill:2002ap} for reviews. The main appeal of technicolor is that we have already encountered similar phenomena in nature: superconductivity is a well known example while relativistic version is the spontaneous chiral symmetry breaking in the vacuum of the ordinary Quantum Chromodynamics (QCD). The earliest models of technicolor were found to have problems with the electroweak (EW) precision data. However, recent developments led to models that have been shown to pass the precision tests \cite{Sannino:2004qp,Dietrich:2005jn}. These models are based on gauge group SU(2) or SU(3) with two Dirac fermions in adjoint or sextet representations, respectively. Phenomenological consequences of these minimal walking technicolor models have been studied. Furthermore, their nonperturbative nature, especially their (near) conformal properties, are being established using lattice simulations, see e.g. \cite{Catterall:2007yx,Hietanen:2008mr,DelDebbio:2008zf,Hietanen:2009az,Bursa:2009we,Shamir:2008pb,DeGrand:2008kx,DeGrand:2010na,Fodor:2009ar,Kogut:2010cz}.

Technicolor predicts the existence of a tower of massive states whose mass is of the order of the EW scale, although pseudo-Goldstone bosons can be lighter. This fact naturally explains why we have not detected technicolor yet. To give masses to the SM fermions one must, however, resort to another unknown sector. Traditional, and most pursued, avenue towards solving for the origin of mass of all matter fields is known as extended technicolor (ETC) \cite{Dimopoulos:1979es}. Here the technicolored matter fermions and standard model matter fields are unified into representations of a larger gauge group, whose breaking is expected to yield the desired Technicolor dynamics as well as the hierarchical Yukawa sector of the Standard Model at low energies. Model building towards this direction is ambitious but difficult, mostly since one does not know for certain even what the desired Technicolor dynamics should be. As alternatives to extended technicolor, one can consider Technicolor coupled to fundamental scalar; this is so-called bosonic technicolor \cite{Simmons:1988fu,Kagan:1991gh,Carone:1992rh,Carone:1994mx,Hemmige:2001vq,Carone:2006wj,Zerwekh:2009yu}. To naturalize this type of models one supersymmetrizes technicolor \cite{Dine:1981za,Dine:1990jd,Dobrescu:1995gz}. Yet another alternative is to complement technicolor with other new strong dynamics which couples to the third generation of QCD quarks. The motivation lies in the heaviness of the top: since its mass is of the order of the electroweak scale, perhaps the top itself plays a role in the electroweak symmetry breaking. The earliest topcolor models \cite{Hill:1991at}, attempted to explain both electroweak symmetry breaking and the generation of top mass itself by condesation of top quarks \cite{Miransky:1988xi, Nambu89, Marciano:1989xd, Bardeen:1989ds}. However, it turned out that the top quark is too light for this scenario to work, and hence Technicolor dynamics was reintroduced to assist in electroweak symmetry breaking, while the light fermion masses are assumed to be due to some unspecified ETC interactions and the top mass itself arises due to the top condensate. This framework generally gives a natural explanation for the large top-bottom mass splitting \cite{Hill:1994hp,Fukano:2008iv,Ryttov:2010fu} in the presence of three QCD quark generations. 

In parallel with the nature of the electroweak symmetry breaking mechanism, another aspect of immediate interest at the upcoming results from the LHC experiment is the generational structure of the Standard Model itself. Especially, whether matter generations beyond the known three exist \cite{Frampton:1999xi}. Recently, there has been some interest in phenomenological models featuring a fourth sequential generation of Standard Model fermions; in particular it has been shown that a heavy fourth generation can be accommodated by the electroweak precision data \cite{Kribs:2007nz}, and the fourth generation quarks help to describe the current experimental data on CP violation and rare decays of B mesons better within the CKM-paradigm \cite{Soni:2010xh}. 

An interesting and less investigated option is the possible existence of non-sequential matter generations, which would arise due to internal consistency of some other BSM sector. An explicit framework, which we consider in this paper, is provided by the minimal walking technicolor (MWT) model \cite{Sannino:2004qp,Dietrich:2005jn}. However, the general features of our results are independent of the underlying Technicolor dynamics and apply to any model with fourth generation matter. The MWT model only provides for the natural existence of new matter generations as explained in \cite{Antipin:2010it}. 

In MWT the gauge group is SU(2)$_{\rm{TC}}\times $SU(3)$_C\times $SU(2)$_L\times$ U(1)$_Y$ and the field content of the technicolor sector is constituted by four Weyl techni-fermions and one techni-gluon, all in the adjoint representation of SU(2)$_{\rm{TC}}$.  The global symmetry of this technicolor theory is SU(4), which breaks explicitly to SU(2)$_L \times $U(1)$_Y$ by the natural choice of the EW embedding \cite{Sannino:2004qp,Dietrich:2005jn}. Electroweak symmetry breaking (EWSB) is triggered by a fermion bilinear condensate, and the vacuum choice is stable against the SM quantum corrections \cite{Dietrich:2009ix}. Taking the electroweak gauge interactions into account, the fermionic particle content of the MWT is represented explicitly by
\beq Q_L^a=\left(\begin{array}{c} U^{a} \\D^{a} \end{array}\right)_L , \qquad U_R^a \
, \quad D_R^a, \quad a=1,2,3. 
\eeq
This particle content suffers from the Witten anomaly \cite{Witten:1982fp}, which is cured by adding an odd number of fermion doublets. These should be singlet under technicolor interaction in order not to spoil the near-conformal behavior, and in order to avoid too large contributions to the $S$-parameter, only choices of one weak doublet (fourth generation of leptons) or three weak doublets (fourth generation of QCD quarks) seem plausible \cite{Antipin:2010it}. We will consider both of these possibilities.


Some possibilities of giving masses to the standard model fermions within MWT have been considered in the literature earlier. An explicit construction of an extended technicolor type model appeared in \cite{Evans:2005pu}. A less natural model introducing a scalar (i.e. bosonic technicolor) mimicking the effects of the extended technicolor interactions has been introduced in \cite{Antola:2009wq}. Supersymmetrized versions of MWT were constructed in \cite{Antola:2010nt}.
In this paper our aim is to  consider the extension of MWT by topcolor dynamics. In order to gain insight into the consequences and constraints of this type of models, we construct two explicit models with different particle content and study their phenomenological implications. 


\section{Model A}


We begin by considering a model based on the MWT in which the SU(2)$_L$ doublet technifermion $Q_L$ has hypercharge $Y(Q_L) = -1/6$ under the U(1)$_Y$ gauge symmetry. In this case, in order to cancel the global and gauge anomalies in the technicolor sector, we consider adding one SM-like SU(2)$_L$ doublet QCD quark \cite{Antipin:2010it}. To explain a large top-bottom splitting, we consider the topcolor-assisted-technicolor (TC2) -type model \cite{Hill:1994hp}. In other words, we extend the ordinary SU(3)$_{\rm QCD} \times $U(1)$_Y$ gauge group to ${\cal G} = $SU(3)$_1 \times $SU(3)$_2 \times $U(1)$_{Y1} \times $U(1)$_{Y2}$ gauge group.
The breaking pattern ${\cal G} \to $SU(3)$_{\rm QCD} \times $U(1)$_Y$ is assumed to occur at some energy $\mu\gg v_{\textrm{weak}}$ leading to the appearance of 8 + 1 massive gauge bosons. We call the eight massive gauge bosons associated with the breaking of SU(3)$_1\times$ SU(3)$_2$ as `` colorons '' and denote these with $G'$. The one massive gauge boson associated with the breaking of U(1)$_1\times$U(1)$_2$ we denote with $Z'$. Their masses are denoted, respectively by $M_{G^\prime}$ and $M_{Z^\prime}$. At low energies, the interactions via the coloron or $Z'$ exchange then lead to effective four fermion interactions which we will write down explicitly below.

\begin{table}[h]
\begin{center}
\begin{tabular}{| c || c | c | c | c | c | c |}
\hline
field & SU(2)$_{\rm TC} $ & SU(3)$_1$  & SU(3)$_2$ & SU(2)$_L$ & U(1)$_{Y1}$ 
& U(1)$_{Y2}$ 
\\
\hline 
$Q_L$ & $\Ysymm$ & 1 & 1 & 2 & 0 & -1/6 
\\
$U_R$ & $\Ysymm$ & 1 & 1 & 1 & 0 & 2/3 
\\
$D_R$ & $\Ysymm$ & 1 & 1 & 1 & 0 & -1/3 
\\ 
\hline
\, & \, & \, & \, & \, & \, & \, 
\\[-2.5ex] 
\hline
$q'_L$ & 1 & 3 & 1 & 2 & 0 & $1/6$ 
\\
$t'_R$ & 1 & 3 & 1 & 1 & 0 & 2/3 
\\ 
$b'_R$ & 1 & 3 & 1 & 1 & 0 & -1/3 
\\ 
\hline
\, & \, & \, & \, & \, & \, & \, 
\\[-2.5ex] 
\hline
$q_L$ & 1 & 3 & 1 & 2 & 1/6 & 0  
\\
$t_R$ & 1 & 3 & 1 & 1 & 2/3 & 0 
\\
$b_R$ & 1 & 3 & 1 & 1 & -1/3 & 0 
\\
\hline 
\, & \, & \, & \, & \, & \, & \, 
\\[-2.5ex] 
\hline 
$l_L$ & 1 & 1 & 1& 2 & -1/2 & 0 
\\ 
$\tau_R$ & 1 & 1 & 1& 1 & -1 & 0 
\\ 
\hline 
\, & \, & \, & \, & \, & \, & \, 
\\[-2.5ex] 
\hline 
$q_i$ & 1 & 1 & 3 & SM & 0 & SM
\\
$l_i$ & 1 & 1 &1 & SM & 0 & SM
\\
\hline
\end{tabular}
\caption{Particle contents and charge assignments of the Model A. The techiquarks are denoted by $Q_L$, $U_R$, $D_R$, while $q_L^\prime$, $t_R^\prime$, $b_R^\prime$ denote the fourth family QCD quarks. The third family quarks/leptons are denoted by $q$ and $l$, respectively. Finally, $q_i/l_i\,\,(i =1,2)$ are the first and second family quarks/leptons and $``{\rm SM}"$ represents the conventional value of the ordinary SM charge.}
\label{particle-model1}
\end{center}
\end{table}%
This minimal model consists of particles in Table \ref{particle-model1}, where the techiquarks are denoted by $Q_L$, $U_R$, $D_R$ and $q_L^\prime$, $t_R^\prime$, $b_R^\prime$ denote the fourth family QCD quarks. The third family quarks/leptons are denoted by $q$ and $l$, respectively. Finally, $q_i/l_i\,\,(i =1,2)$ are the first and second family quarks/leptons and $``{\rm SM}"$ represents the conventional value of the ordinary SM charge. We consider the case that SU(3)$_1$ gauge coupling $(h_1)$ is stronger than SU(3)$_2$ gauge coupling $(h_2)$, i.e. the ratio $\cot \theta = h_1/h_2$ is larger than one. Furthermore, we consider the U(1)$_{Y1}$ gauge coupling $(h'_1)$ to be stronger than the U(1)$_{Y2}$ gauge coupling $(h'_2)$,  which implies that the ratio $\cot \theta' = h'_1/h'_2 $ is larger than one. In the notation introduced above, the QCD coupling and hypercharge U(1)$_Y$ couplings are given by  $g_{\rm QCD} = h_1 \sin \theta = h_2 \cos \theta$ and $g_Y = h'_1 \sin \theta' = h'_2 \cos \theta'$.

Let us now write explicitly the model Lagrangian at energies below the scale $\mu$, i.e. after the breaking ${\cal G} \to SU(3)_{\rm QCD} \times U(1)_Y$.  Then we can divide the model into three parts, which are the SM part, the MWT part and the four fermion interaction part. We concentrate on the four fermion sector and we neglect the first and second family fermions for simplicity; this is justified since the masses of the light fermions are neglected (or assumed to be explained by some underlying ETC dynamics operating at much higher scales) and we aim to explain only the mass patterns of the third and fourth quark generations here. The coloron exchange gives 
\beq
{\cal L}^{4f}_C =
-\frac{4\pi \kappa_3}{M^2_{G'}} \times 
\left[ 
\begin{aligned}
& \bar{q'}_L \gamma_\mu T^a q'_L + \bar{t'}_R \gamma_\mu T^a t'_R + \bar{b'}_R \gamma_\mu T^a b'_R \\[1ex]
&\quad + \bar{q}_L \gamma_\mu T^a q_L + \bar{t}_R \gamma_\mu T^a t_R + \bar{b}_R \gamma_\mu T^a b_R 
\end{aligned}
\right]^2\,,
\eeq
where $\kappa_3 \equiv \alpha_{\rm QCD} \cot^2 \theta$ and $T^a$ is the Gell-Mann matrix. The $Z'$ exchange gives 
\beq
{\cal L}^{4f}_{Z'} &=&
-\frac{4\pi \kappa_1}{M^2_{Z'}} 
\left[ \frac{1}{6} \bar{q}_L \gamma_\mu q_L + \frac{2}{3} \bar{t}_R \gamma_\mu t_R 
            - \frac{1}{3} \bar{b}_R \gamma_\mu b_R 
            -\frac{1}{2} \bar{l}_L \gamma_\mu l_L - \bar{\tau}_R \gamma_\mu \tau_R
\right]^2 \nonumber \\
&&
-\frac{4\pi}{M^2_{Z'}} \frac{\alpha^2_Y}{\kappa_1}
\left[ 
\begin{aligned}
&\frac{1}{6}\bar{q'}_L \gamma_\mu q'_L + \frac{2}{3}\bar{t'}_R \gamma_\mu t'_R  - \frac{1}{3}\bar{b}_R \gamma_\mu b'_R \\[1ex]
&\quad - \frac{1}{6} \bar{Q}_L \gamma_\mu Q_L + \frac{2}{3} \bar{U}_R \gamma_\mu U_R  - \frac{1}{3} \bar{D}_R \gamma_\mu D_R
\end{aligned}
\right]^2
\label{4fZprime-modelB}
\,,
\eeq
where $\kappa_1 \equiv \alpha_Y \cot^2 \theta'$. 
After the Fierz rearrangement these vector-vector four fermion interactions can be written as
\beq
{\cal L}^{4f}_{\rm model A}
=
&& \hspace*{-3ex}
\frac{\pi \kappa_3}{M^2_{G'}} 
\left[ 
\begin{aligned}
&(\bar{q'}_L  t'_R)^2 + (\bar{q'}_L  b'_R)^2 + (\bar{q}_L  t_R)^2 + (\bar{q}_L  b_R)^2 \\
& \quad +(\bar{q'}_L  t_R)^2 + (\bar{q'}_L  b_R)^2 + (\bar{q}_L  t'_R)^2 + (\bar{q}_L  b'_R)^2+\cdots
\end{aligned}
\right]
\nonumber\\
&&
+
\frac{2\pi \kappa_1}{M^2_{Z'}} 
\left[ \frac{1}{9}(\bar{q}_L  t_R)^2 - \frac{1}{18}(\bar{q}_L  b_R)^2 + \frac{1}{2} (\bar{l}_L \tau_R)^2 \right]
\label{4fermi-model1}\\
&&
+
\frac{2\pi}{M^2_{Z'}} \frac{\alpha^2_Y}{\kappa_1}
\left[ \frac{1}{9}(\bar{q'}_L  t'_R)^2 - \frac{1}{18}(\bar{q'}_L  b'_R)^2  - \frac{1}{9} (\bar{Q}_L U_R)^2 + \frac{1}{18} (\bar{Q}_L D_R)^2\right]
\nonumber
\,.
\eeq
Here we have not explicitly written quark-lepton type four fermion interactions; these are proportional to $\kappa_1/M_{Z^\prime}^2$ and parametrically small in comparison to the interactions proportional to $\kappa_3$.

Next, we turn to the phenomenological analysis of this model.  We treat the four fermion interactions detailed above with the gap equation of the NJL model which is sufficient in this case. This can be justified through the following argument: The technifermions have TC charges and their gauge interactions are strong and trigger the technifermion condensation. The four fermion interactions among technifermions in Eq.(\ref{4fermi-model1}) are proportional to $\alpha^2_Y/\kappa_1$ and hence negligible compared with $\kappa_{3,1}$ terms in Eq.(\ref{4fermi-model1}). Thus the EWSB by the technifermions will be triggered almost entirely by the TC gauge dynamics. On the other hand, after ${\cal G} \to$ SU(3)$_{\rm QCD} \times $U(1)$_Y$, the SM-fermions and fourth family fermions have ordinary SM charges only, and the SM gauge couplings at the EW scale are too small to trigger the EWSB. Hence the condensation of the SM and fourth family fermions is controlled dominantly by the four fermion interactions in Eq.(\ref{4fermi-model1}), and it is enough to use the NJL (instead of the gauged-NJL) analysis for the criticality of the EWSB by the SM-fermions or fourth family fermion condensations. The gap equation in the NJL model is given by
\beq
1 = G  \times \frac{2 N \Lambda^2}{4\pi^2} \left[ 1 - \frac{m^2_{\rm dyn}}{\Lambda^2} \ln \frac{\Lambda^2}{m^2_{\rm dyn}}\right]\,,
\eeq
where $m_{\rm dyn}$ is the dynamical mass of fermions and $N = N_c$ for quarks, and $N=1$ for leptons.
The ultraviolet scale below which the NJL model is defined is denoted by $\Lambda$, and we set $\Lambda \sim M_{G'} \sim M_{Z'}$ in this paper. If $2 N \Lambda^2 G/(4\pi^2) > g_{\rm crit} =1$, fermions obtain a dynamical mass, $m_{\rm dyn} \neq 0$.
For simplicity we define the dimensionless four fermion coupling $g \equiv 2 N M^2 G/(4\pi^2)$ for each channel in model A as
\beq
&&
g_{t'} = \frac{N_c}{2\pi} \left[ \kappa_3 + \frac{2}{9 N_c} \frac{\alpha^2_Y}{\kappa_1} \right]
\quad , \quad
g_{b'} = \frac{N_c}{2\pi} \left[ \kappa_3 - \frac{1}{9 N_c} \frac{\alpha^2_Y}{\kappa_1} \right]
\label{g-model1-4th}
\,,\\[1ex]
&&
g_{t} = \frac{N_c}{2\pi} \left[ \kappa_3 + \frac{2}{9 N_c} \kappa_1 \right]
\quad ,\quad
g_{b} = \frac{N_c}{2\pi} \left[ \kappa_3 - \frac{1}{9 N_c} \kappa_1 \right]
\label{g-model1-tb}
\,,\\[1ex]
&&
g_\tau = \frac{1}{2\pi} \kappa_1\,,
\label{g-model1-tau}
\,\\[1ex]
&&
g_{t_Lt'_R} = g_{t'_Lt_R} = g_{b_Lb'_R} = g_{b'_Lb_R} =  \frac{N_c}{2\pi} \kappa_3
\label{g-model1-34mix}
\,.
\eeq
As we already discussed, the technifermion condensation is dominantly triggered by the TC gauge interaction since $g_{U,D} \propto \alpha^2_Y/\kappa_1 \ll 1$. On the other hand, $t',b',t,b$-condensations will be triggered if the conditions
\beq
g_{t',b',t,b} > g_{\rm crit} = 1,
\label{crit-cond-model1}
\eeq
are satisfied. In addition we require $g_{\tau} < g_{\rm crit} = 1$ to avoid condensation in $\tau$-channel. Note that we allow for the formation of $b$-condensate. This is essential for the explanation of physical masses of third and fourth generation quarks as we will show below. Note that this is the main difference in comparison to the models with three QCD generations only, since then one needs to assume that $b$-quarks do not condense. 
If the conditions in Eq.(\ref{crit-cond-model1}) are satisfied, we may obtain additional condensation: $g_{t_Lt'_R} = g_{t'_Lt_R} = g_{b_Lb'_R} = g_{b'_Lb_R} > g_\crit $ since $g_{t'} > g_{t_Lt'_R} > g_{b'}$ as one can see from Eqs. (\ref{g-model1-4th}) and (\ref{g-model1-34mix}). 
Thus after the condensates are formed the mass terms of $t,t',b,b'$ arising from (\ref{4fermi-model1}) are
\beq
{\cal L}_{\rm mass} =
-\bpm \bar{t}_L & \bar{t'}_L\epm \!\!
\bpm m_0^t & m^{tt'}_{\rm dyn} \\[1ex] m^{tt'}_{\rm dyn} & M^{t'}_{\rm dyn} \epm \!\!
\bpm t_R \\[1ex] t'_R\epm
-\bpm \bar{b}_L & \bar{b'}_L\epm \!\!
\bpm m_0^b & m^{bb'}_{\rm dyn} \\[1ex] m^{bb'}_{\rm dyn} & M^{b'}_{\rm dyn} \epm \!\!
\bpm b_R \\[1ex] b'_R\epm 
+ \text{h.c.}
\,.
\label{masstextures}
\eeq
Due to Eq. (\ref{g-model1-4th}) with $\alpha^2_Y/\kappa_1 \ll \kappa_3$, we set  $M^{t'}_{\rm dyn} \simeq M^{b'}_{\rm dyn} \equiv M_0$, and furthermore  $m^{tt'}_{\rm dyn} = m^{bb'}_{\rm dyn} \simeq M_0$ due to Eqs. (\ref{g-model1-4th}) and (\ref{g-model1-34mix}). So ${\cal L}_{\rm mass}$ can be rewritten as
\beq
{\cal L}_{\rm mass} &\simeq&
-\bpm \bar{t}_L & \bar{t'}_L\epm \!\!
\bpm m_0^t & M_0 \\[1ex] M_0 & M_0 \epm \!\!
\bpm t_R \\[1ex] t'_R\epm
-\bpm \bar{b}_L & \bar{b'}_L\epm \!\!
\bpm m_0^b & M_0 \\[1ex] M_0 & M_0 \epm \!\!
\bpm b_R \\[1ex] b'_R\epm 
+ \text{h.c.} \,\nonumber\\
&=&
-\bpm \bar{t}_L & \bar{t'}_L\epm \!
{\cal M}_t \!
\bpm t_R \\[1ex] t'_R\epm
-\bpm \bar{b}_L & \bar{b'}_L\epm \!
{\cal M}_b \!
\bpm b_R \\[1ex] b'_R\epm 
+ \text{h.c.}
\,.
\eeq
To obtain the (squared) masses of the physical states, we diagonalize the matrices ${\cal M}_{t(b)}{\cal M}^T_{t(b)}$, and as a result we obtain the eigenvalues as
\beq
m^2_t(\pm) &=& 
\frac{1}{2} \left[ (m_0^t)^2 + 3 M^2_0 \pm \sqrt{((m^t_0)^2 + M^2_0 )^2 + 4M^2_0(M^2_0+2M_0 m^t_0)} 
\right]
\,,\label{topmass}\\
m^2_b(\pm) &=& \frac{1}{2} \left[ ((m_0^b)^2 + 3 M^2_0 \pm \sqrt{(m^b_0)^2 + M^2_0 )^2 + 4M^2_0(M^2_0+2M_0 m^b_0)} 
\right]
 \,.\label{bottommass}
\eeq 
The masses $m_t(-), m_t(+), m_b(-), m_b(+),$ are then identified, respectively, as top quark mass ($m_t$), $t'$ quark mass ($m_{t'}$), bottom quark mass ($m_b$) and $b'$ quark mass ($m_{b'}$). The above mass matrices allow for several different phenomenological mass textures. The bottom quark can be light in comparison to other third and fourth generation quarks provided $m_0^b\simeq M_0$. Then, the limit $m_{b'} > 338 \GeV$ \cite{Aaltonen:2009nr} restricts $M_0\ge 176$ GeV. As an example, we can take $M_0=400$ GeV, and then the condition to obtain top quark mass $176$ GeV from Eq. (\ref{topmass}) gives $m_0^t=474$ GeV, and results in the mass value 821 GeV for the $t^\prime$ quark. Note that from the strong dynamics viewpoint the ``democratic'' structure of the mass matrices in Eq. (\ref{masstextures}) is very natural and leads to mass patterns compatible with observations.

In addition, the condensates ensuing the above dynamical masses also contribute to EWSB. Generally we have
\begin{equation}
v_{\rm{weak}}^2=f_{\rm{TC}}^2+f_{tt}^2+f_{bb}^2+f_{tt^\prime}^2+f_{bb^\prime}^2+f_{t^\prime t^\prime}^2+f_{b^\prime b^\prime}^2,
\label{ewscale}
\end{equation} 
where $v_{\rm{weak}}=246$ GeV and the ``decay constant'' $f_i$ is related to the corresponding dynamical mass as
\begin{equation}
f_i^2=\frac{N_c}{8\pi^2}m_{{\rm{dyn}},i}^2\ln\left(\frac{\Lambda^2}{m^2_{{\rm{dyn}},i}}\right).
\label{fiequ}
\end{equation}
For an estimate, assume all $f_i$ equal, and then (\ref{ewscale}) implies $f_i\sim 90$ GeV, which then in turn from (\ref{fiequ}) implies
\begin{equation}
m_{{\rm{dyn}},i}^2\sim \sqrt{\frac{8\pi^2}{N_c}}f_i^2 \sim 480{\textrm{~~GeV}},
\end{equation}
consistently with the numbers we have quoted above. Clearly the fact that several condensates contribute to EWSB leads to constrain the dynamical masses rather close to the current observational limits. This makes it possible to exclude this model at the LHC if no fourth generation quarks will be observed. Finally we note that this model is not at odds with the current EW precision data \cite{Antipin:2010it}.



\section{Model B}

Next, we consider a model based on the MWT in which the SU(2)$_L$ doublet technifermion $(Q_L)$ has $Y(Q_L) = 1/6$ under the U(1)$_Y$ gauge symmetry. In this case, in order to cancel the gauge anomaly in the technicolor sector, we should add one SM-like SU(2)$_L$ doublet of leptons \cite{Dietrich:2005jn}. Moreover, we would again like to explain the large top-bottom splitting, and we do this in the framework of the top condensation. Model B is based on the top quark seesaw model \cite{Chivukula:1998wd}. In this model, among the SM quarks, the third family quarks obtain their masses only by the seesaw mechanism after some condensations are triggered, but other quarks obtain their masses mainly by ETC interactions with the technifermion condensates.

Along this line, a minimal model consists of particles in Table \ref{particle-model2} where $q/l$ are the SM-third family quarks/leptons, $q_i/l_i\,\,(i =1,2)$ are the SM- first and second family quarks/leptons and ``SM" represents the ordinary SM charge values. We note that $t',b'$ are vector-like fermions under the EW gauge symmetry. As in model A, we extend the ordinary SU(3)$_{\rm QCD} \times $U(1)$_Y$ gauge group to ${\cal G} = $SU(3)$_1 \times $SU(3)$_2 \times $U(1)$_{Y1} \times $U(1)$_{Y2}$ which we assume to break according to the pattern ${\cal G} \to $SU(3)$_{\rm QCD} \times $U(1)$_Y$.
\begin{table}[h]
\begin{center}
\begin{tabular}{| c || c | c | c | c | c | c |}
\hline
field & SU(2)$_{\rm TC} $ & SU(3)$_1$  & SU(3)$_2$ & SU(2)$_L$ & U(1)$_{Y1}$ 
& U(1)$_{Y2}$ 
\\
\hline 
$Q_L$ & $\Ysymm$ & 1 & 1 & 2 & 0 & 1/6 
\\
$U_R$ & $\Ysymm$ & 1 & 1 & 1 & 0 & 2/3 
\\
$D_R$ & $\Ysymm$ & 1 & 1 & 1 & 0 & -1/3 
\\ 
\hline
\, & \, & \, & \, & \, & \, & \, 
\\[-2.5ex] 
\hline
$L_L$ & 1 & 1 & 1 & 2 & 0 & -1/2
\\
$N_R$ & 1 & 1 & 1 & 1 & 0 & 0
\\ 
$E_R$ & 1 & 1 & 1 & 1 & 0 & -1 
\\ 
\hline
\, & \, & \, & \, & \, & \, & \, 
\\[-2.5ex] 
\hline
$q_L$ & 1 & 3 & 1 & 2 & 1/6 & 0  
\\
$t_R$ & 1 & 1 & 3 & 1 & 2/3 & 0 
\\
$b_R$ & 1 & 1 & 3 & 1 & -1/3 & 0 
\\
\hline 
\, & \, & \, & \, & \, & \, & \, 
\\[-2.5ex] 
\hline 
$l_L$ & 1 & 1 & 1& 2 & -1/2 & 0 
\\ 
$\tau_R$ & 1 & 1 & 1& 1 & -1 & 0 
\\ 
\hline 
\, & \, & \, & \, & \, & \, & \, 
\\[-2.5ex]
\hline
$t'_L$ & 1 & 1 & 3 & 1 & 2/3 & 0  
\\
$t'_R$ & 1 & 3 & 1 & 1 & 2/3 & 0 
\\
$b'_L$ & 1 & 1 & 3 & 1 & -1/3 & 0 
\\
$b'_R$ & 1 & 3 & 1 & 1 & -1/3 & 0 
\\
\hline 
\, & \, & \, & \, & \, & \, & \, 
\\[-2.5ex] 
\hline 
$q_i$ & 1 & 1 & 3 & SM & 0 & SM
\\
$l_i$ & 1 & 1 &1 & SM & 0 & SM
\\
\hline
\end{tabular}
\caption{Particle content and charge assignments of the Model B.  Here $q/l$ are the SM-third family quarks/leptons, $q_i/l_i\,\,(i =1,2)$ are the SM- first and second family quarks/leptons and ``SM'' represents the ordinary SM charge.} 
\label{particle-model2}
\end{center}
\end{table}%

After this symmetry breaking we can again divide this model into three parts:
the SM part, the MWT part and  the four fermion interaction part.
It is again sufficient to concentrate on the four fermion sector and also to neglect the first and second family fermions.
The coloron, with mass $M_{G'}$, exchange gives 
\beq
{\cal L}^{4f}_C &=&
-\frac{4\pi \kappa_3}{M^2_{G'}} 
\left[ 
\bar{q}_L \gamma_\mu T^a q_L + \bar{t'}_R \gamma_\mu T^a t'_R + \bar{b'}_R \gamma_\mu T^a b'_R 
\right] 
\nonumber\\
&&
-\frac{4\pi}{M^2_{G'}} \frac{\alpha^2_{\rm QCD}}{\kappa_3}
\left[ 
\bar{t}_R \gamma_\mu T^a t_R + \bar{b}_R \gamma_\mu T^a b_R +
\bar{t'}_L \gamma_\mu T^a t'_L + \bar{b'}_L \gamma_\mu T^a b'_L 
\right]^2\,,
\eeq
and the $Z'$, whose mass is $M_{Z'}$, exchange gives 
\beq
{\cal L}^{4f}_{Z'} &=&
-\frac{4\pi \kappa_1}{M^2_{Z'}} \times 
\left[ 
\begin{aligned}
& \frac{1}{6}\bar{q}_L \gamma_\mu q_L + \frac{2}{3} \bar{t}_R \gamma_\mu t_R -\frac{1}{3} \bar{b}_R \gamma_\mu b_R \\[1ex]
&\quad + \frac{2}{3} \bar{t'}_L \gamma_\mu  t'_L + \frac{2}{3}\bar{t'}_R \gamma_\mu t'_R 
- \frac{1}{3} \bar{b'}_L \gamma_\mu  b'_L - \frac{1}{3}\bar{b'}_R \gamma_\mu b'_R  \\[1ex] 
& \quad \quad -\frac{1}{2} \bar{l}_L \gamma_\mu l_L - \bar{\tau}_R \gamma_\mu \tau_R+\cdots \\[1ex]
\end{aligned}
\right]^2\,,
\label{4fermi-model2-Zprime}
\eeq
where $\kappa_3 \equiv \alpha_{\rm QCD} \cot^2 \theta$ and $\kappa_1 \equiv \alpha_Y \cot^2 \theta'$. 
After the Fierz rearrangement these vector-vector four fermion interactions can be written as
\beq
{\cal L}^{4f}_{\rm model B}
=
&& \hspace*{-2ex}
\frac{\pi \kappa_3}{M^2_{G'}} \left[ (\bar{q}_L  t'_R)^2 + (\bar{q}_L  b'_R)^2 \right]
+ \frac{\pi}{M^2_{G'}} \frac{\alpha^2_{\rm QCD}}{\kappa_3}
\left[ 
(\bar{t'}_L  t_R)^2 + (\bar{b'}_L  t_R)^2 +(\bar{t'}_L  b_R)^2 + (\bar{q'}_L  b_R)^2 
\right]
\nonumber\\[1ex]
&&
+
\frac{2\pi \kappa_1}{M^2_{Z'}} 
\left[ 
\begin{aligned}
&\frac{1}{9}(\bar{q}_L  t_R)^2 + \frac{1}{9}(\bar{q}_L  t'_R)^2 
- \frac{1}{18}(\bar{q}_L  b_R)^2 - \frac{1}{18}(\bar{q}_L  b'_R)^2 + \frac{1}{2} (\bar{l}_L \tau_R)^2 
\label{4fermi-model2}
\\[1ex]
& \quad \quad 
+ \frac{4}{9}(\bar{t'}_L  t_R)^2 + \frac{4}{9}(\bar{t'}_L  t'_R)^2 
+ \frac{1}{9}(\bar{b'}_L  b_R)^2  +\frac{1}{9} (\bar{b'}_L b'_R)^2 +\cdots
\end{aligned}
\right]
\,,
\eeq
where we again neglect quark-lepton type four fermion interactions; the most important terms will anyway be the ones proportional to $\kappa_3$. 

The essential point, which leads to the desired seesaw mechanism for the quark masses is, that in this model, after ${\cal G} \to $SU(3)$_{\rm QCD} \times $U(1)$_Y$ symmetry breaking, we are allowed to add SM gauge invariant mass terms 
\begin{equation}
{\cal L}^0_{\rm mass} 
=
-\mu_t \bar{t}^\prime_Lt_R-M_{t^\prime}\bar{t}^\prime_Lt_R^\prime-\mu_b\bar{b}^\prime_Lb_R-M_{b^\prime}\bar{b}^\prime_L b^\prime_R+ {\rm h.c.}\,
\label{baremass-model2}
\end{equation}
Since our approach is bottom-up model building, the origin of these terms is left unspecified. For example, the $M_{t^\prime}$ and $M_{b^\prime}$ terms could arise from the vacuum expectation value of a scalar field in $(\bar{3},3)$ representation of SU(3)$_1\times$ SU(3)$_2$, while the $\mu_t$ and $\mu_b$ terms could be included by some underlying ETC interactions operating at yet higher scales in comparison to $M_{G^\prime}$ and $M_{Z^\prime}$ relevant for the breaking of SU(3)$_1\times$SU(3)$_2$. The different origin of these contributions will be reflected at low energies as hierarchical structure $\mu_t\ll M_{t^\prime}$ and $\mu_b\ll M_{b^\prime}$.

Give the above four-fermion interactions and bare mass terms, we then turn to the model dynamics. As in the case of model A considered in previous section, we again apply the NJL model analysis. We can define the dimensionless four fermion couplings for each channel in model B as
\beq
&&
g_{t'} = \frac{N_c}{2\pi} \frac{8}{9 N_c} \kappa_1
\quad ,\quad
g_{b'} = \frac{N_c}{2\pi} \frac{2}{9 N_c} \kappa_1 
\quad , \quad
g_{t} = \frac{N_c}{2\pi} \frac{2}{9 N_c} \kappa_1 
\quad ,\quad
g_{b} = -\frac{N_c}{2\pi} \frac{1}{9 N_c} \kappa_1 
\,,\\[1ex]
&&
g_\tau = \frac{1}{2\pi} \kappa_1\,,
\quad ,\quad
g_{t_Lt'_R} = \frac{N_c}{2\pi} \left[ \kappa_3 + \frac{2}{9 N_c} \kappa_1 \right]
\quad , \quad
g_{b_Lb'_R} = \frac{N_c}{2\pi} \left[ \kappa_3 - \frac{1}{9 N_c} \kappa_1 \right]
\label{g-model2-34mix1}
\,,\\[1ex]
&&
g_{t'_Lt_R} = \frac{N_c}{2\pi} \left[ \frac{\alpha^2_{\rm QCD}}{\kappa_3} + \frac{8}{9 N_c} \kappa_1 \right]
\quad , \quad
g_{b'_Lb_R} = \frac{N_c}{2\pi} \left[ \frac{\alpha^2_{\rm QCD}}{\kappa_3} + \frac{2}{9 N_c} \kappa_1  \right]
\label{g-model2-34mix2}
\eeq
where we again assume $\Lambda \sim M_{G'} \sim M_{Z'}$.
Similarly as in model A, the technifermion condensation is triggered dominantly by the TC gauge interaction since $g_{U,D} \propto \alpha^2_Y/\kappa_1 \ll 1$, and this justifies the use on NJL model analysis only.
In this model, the electroweak symmetry breaking is triggered by the condensates $\vev{\bar{t}_L t'_R} \neq 0$ and $\vev{\bar{b}_L b'_R} \neq 0$. These condensates are formed when the criticality conditions
\beq
g_{t_Lt'_R}\, , \,  g_{b_Lb'_R}> g_{\rm crit} = 1 \quad ,\quad 
g_{t',b't,b,\tau} \, , \,  g_{t'_Lt_R} \, , \,g_{b'_Lb_R}< g_{\rm crit} = 1\,,
\label{crit-cond-model2}
\eeq
are satisfied. Consequently, this leads to the minimal top seesaw model \cite{Chivukula:1998wd} for the sector of the third and fourth families.

After the condensates are formed, the mass terms of $t,t',b,b'$ arising from Eqs. (\ref{4fermi-model2},\ref{baremass-model2}) are
\beq
{\cal L}_{\rm mass} =
-\bpm \bar{t}_L & \bar{t'}_L\epm \!\!
\bpm 0 & m^{tt'}_{\rm dyn} \\[1ex] \mu_t & M_{t'} \epm \!\!
\bpm t_R \\[1ex] t'_R\epm
-\bpm \bar{b}_L & \bar{b'}_L\epm \!\!
\bpm 0 & m^{bb'}_{\rm dyn} \\[1ex] \mu_b & M_{b'} \epm \!\!
\bpm b_R \\[1ex] b'_R\epm 
+ \text{h.c.}
\,,
\eeq
where $m^{t'}_{\rm dyn} \simeq m^{b'}_{\rm dyn} \equiv M_0$ because of Eq.(\ref{g-model2-34mix1}) with $\alpha^2_Y/\kappa_1 \ll \kappa_3$,.
Hence, ${\cal L}_{\rm mass}$ can be rewritten as \cite{Chivukula:1998wd}
\beq
{\cal L}_{\rm mass} &\simeq&
-\bpm \bar{t}_L & \bar{t'}_L\epm \!\!
\bpm 0 & M_0 \\[1ex] \mu_t & M_{t'} \epm \!\!
\bpm t_R \\[1ex] t'_R\epm
-\bpm \bar{b}_L & \bar{b'}_L\epm \!\!
\bpm 0 & M_0 \\[1ex] \mu_b & M_{b'} \epm \!\!
\bpm b_R \\[1ex] b'_R\epm 
+ \text{h.c.} \,\nonumber\\
&=&
-\bpm \bar{t}_L & \bar{t'}_L\epm \!
{\cal M}_t \!
\bpm t_R \\[1ex] t'_R\epm
-\bpm \bar{b}_L & \bar{b'}_L\epm \!
{\cal M}_b \!
\bpm b_R \\[1ex] b'_R\epm 
+ \text{h.c.}
\,.
\eeq


It is straightforward to obtain $t,t^\prime$ and $b,b^\prime$ masses by diagonalizing the above mass matrices. For simplicity, we furthermore assume the the mass matrices are of the seesaw-type, i.e. the scales $M_{t^\prime}$ and $M_{b^\prime}$ dominate over the scales $\mu_t$, $\mu_b$ and $M_0$. This hierarchical structure is dependent on the underlying dynamics which we have not specified. However, as we discussed below Eq. (\ref{baremass-model2}), plausible possibilities resulting in this type of structures can be easily imagined.

With the assumed hierarchy, then, the top quark mass is approximately given by
\begin{equation}
m_t\simeq M_0\left(\frac{\mu_t}{M_{t^\prime}}\right)\left[1-\frac{M_0^2+\mu_t^2}{2M_{t^\prime}^2}\right],
\end{equation}
and a similar formula with the replacement of $t$ with $b$ and $t^\prime$ with $b^\prime$ holds for the $b$-quark mass. In particular, the ratio of top and bottom quark masses is approximately
\begin{displaymath}
\frac{m_t}{m_b}\sim
\left(\frac{\mu_t}{\mu_b}\right)\left(\frac{M_b^\prime}{M_t^\prime}\right).
\end{displaymath}
This equation shows that in this model the large observed hierarchy between the top and bottom quark masses can be explained by more modest hierarchies between the different scales in $b,b^\prime$ and $t,t^\prime$ sectors. 

In addition to the mass patterns of $t$ and $b$ quarks, several other constraints should be taken into consideration \cite{Popovic:1998vb}. In this paper, we will consider two constraints: First, $\Delta \rho_\ast$ should be compatible with the existing precision data. Second, the Landau pole of the coupling of U(1)$_{Y_1}$ gauge group (which is stronger of the two U(1) factors) should be sufficiently far above the symmetry breaking scale $\Lambda$. 

First we consider constraint from the shift in the ratio of $W$ and $Z$ boson masses due to beyond the Standard Model physics contributions, i.e. $\Delta \rho_\ast$ which is related to the $T$ parameter and defined as
\begin{equation}
\Delta\rho_\ast\equiv \alpha T=\frac{e^2}{s_W c_WM_Z^2}\left(\Pi_{11}(0)-\Pi_{33}(0)\right).
\end{equation}
Here $\Pi_{ii}$ denotes the contributions proportional to $ig_{\mu\nu}$ in the $W$ and $Z$ vacuum polarizations with the gauge couplings factored out, $e$ is the electromagnetic gauge coupling, $s_W$ and $c_W$ denote the usual weak mixing angle and $M_Z$ the $Z$ boson mass.
 
Explicitly, in model B, $\Delta \rho*$ is divided as 
\beq
\Delta \rho_* = 
\Delta \rho_*^{\rm TC} + \Delta \rho_*^{\rm L} 
+ \Delta \rho_*^C + \Delta \rho_*^{Z'} +  \Delta \rho_*^{t',b'} \,,
\eeq
where each term,  from left to right, in the r.h.s represents a contribution from the technicolor sector, the new lepton sector, colorons, $Z'$ and new vector like quarks, respectively.
$\Delta \rho_*^{\rm TC}$ and $\Delta \rho_*^{\rm L}$ are proportional to $\alpha^2_Y/\kappa_1$ and hence small thanks to our charge assignments which differs from $\Delta \rho_*^{\rm TC}$ in \cite{Chivukula:1995dc}. 
Obtaining $\Delta \rho_*^{t',b'}$ is complicated due to mixing between $t (b)$ and $t' (b')$. To obtain and initial estimate, we neglect $\Delta \rho_*^{t',b'}$ and consider the upper bound for $\Delta \rho_*^C + \Delta \rho_*^{Z'}$. This will then provide a rough constraint as
\beq
\Delta \rho_*^C + \Delta \rho_*^{Z'} < \left[ \Delta \rho_* \right]_{\rm exp.}\,,
\label{const-delta-rho}
\eeq
which is the same as the constraint for $\Delta \rho_*$ in \cite{Popovic:1998vb}.
The experimental bound of $\Delta \rho_*$ ($\left[ \Delta \rho_* \right]_{\rm exp.}$) is 
$\left[ \Delta \rho_* \right]_{\rm exp.} < 0.4 \% \,\,\, (95 \% \cl)$ \cite{Chivukula:1995dc}.

In model B $\Delta \rho_*^C$  is evaluated within the NJL approximation \cite{Chivukula:1995dc} and is given by 
\beq
\Delta \rho_*^C = \frac{16 \pi^2 \alpha_Y}{3 s^2_W c^2_W} \left( \frac{f^2_t}{M_Z M_{G'}} \right)^2 \times \kappa_3\,.
\label{delta-rho-C}
\eeq
Here $f_t$ is the ``decay constant'' for the top quark condensate defined in Eq. (\ref{fiequ})
with $m_{\rm dyn}$ the dynamical mass arising due to 
formation of $\bar{t_L} t'_R$ condensate. In addition, in model B, there is another dynamical mass term, $m'_{\rm dyn} \bar{b_L} b'_R$ arising from the formation of $\vev{\bar{b_L} b'_R} \neq 0$. Now, for simplicity, we assume $m_{\rm dyn} \simeq m'_{\rm dyn}$ which implies $g_{t_Lt'_R} \simeq g_{b_Lb'_R}$ and this is correct if $\kappa_1$ is not large. This is turns out to be the case since, first, $\kappa_1$ is bounded from above in any case to prevent the formation of $\tau$ condensate and, second, even more stringent upper bound will be provided from the constraint on the Landau pole. Hence, the vev ($v_{\rm EW} = 246 \GeV$) for the EWSB is given by 
\beq
v^2_{\rm EW} = f^2_{\rm TC} + f^2_t + f^2_b \simeq  f^2_{\rm TC} + 2 f^2_t\,.
\eeq
On the other hand, $\Delta \rho_*^{Z'}$, which comes from the $Z-Z'$ mixing \cite{Chivukula:1996cc}, is given by
\beq
\Delta \rho_*^{Z'} =
\frac{\alpha_Y s^2_W}{\kappa_1} \frac{M^2_Z}{M^2_{Z'}} 
\left[ 1 + \left( \epsilon_t + \epsilon_b \right) \frac{\alpha_Y + \kappa_1}{\alpha_Y} \right]\,,
\label{delta-rho_Zp-0}
\eeq
where $\epsilon_t = 2 \left( f^2_t/v^2_{\rm EW} \right) \left( Y^{t_L}_1- Y^{t'_R}_1 \right)$ and $\epsilon_b = 2 \left( f^2_b/v^2_{\rm EW} \right) \left( Y^{b_L}_1- Y^{b'_R}_1 \right)$ with $f^2_b \simeq f^2_t$.
Note that the result in Eq.(\ref{delta-rho_Zp-0}) is different from $\Delta \rho_*^{Z'}$ in \cite{Chivukula:1996cc}. This is so since in \cite{Chivukula:1996cc} only $\vev{\bar{t}_L t_R} \neq 0$, which corresponds to $\vev{\bar{t}_L t'_R} \neq 0$ in model B, is formed so the case in \cite{Chivukula:1996cc} corresponds to $\epsilon_b = 0$ in Eq.(\ref{delta-rho_Zp-0}). 
On the other hand In model B $- \epsilon_t = \epsilon_b = 1/2$ so in our model $\Delta \rho_*^{Z'}$ is given by 
\beq
\Delta \rho_*^{Z'} = \frac{\alpha_Y s^2_W}{\kappa_1} \frac{M^2_Z}{M^2_{Z'}} \,.
\label{delta-rho_Zp}
\eeq

Combining these results, Eq.(\ref{const-delta-rho}) becomes 
\beq
\frac{16 \pi^2 \alpha_Y}{3 s^2_W c^2_W} \left( \frac{f^2_t}{M_Z M_{G'}} \right)^2 \times \kappa_3
+
\alpha_Y s^2_W \frac{M^2_Z}{M^2_{Z'}} \times \frac{1}{\kappa_1}
< 0.4 \%\,.
\label{const-delta-final}
\eeq
For example, consider the modest hierarchy $M_{G^\prime}\approx M_{Z^\prime}\approx 10 M_Z$ and $f_t\sim M_Z$. Then the second term on the left-hand side of Eq. (\ref{const-delta-final}) is $\sim 10^{-4}\%$ for $\kappa_1\sim{\cal{O}}(1)$ and negligible to the first term which, under the same approximations, is $\sim 10^{-1}\%$. Hence, this constraint provides a weak bound on $\kappa_3$ and, as we show below, will be superseded by the other constraints.  

Next, we consider constraint from a position of the Landau pole of the $U(1)_{Y_1}$ coupling which we assume to be more strongly coupled of the two $U(1)$ factors. The running of $\alpha_{Y1}$ at one-loop order is 
\beq
\frac{1}{\alpha_{Y1}(\mu)} - \frac{1}{\alpha_{Y1}(\Lambda_{\rm UV})} = \frac{b_{Y1}}{2\pi} \ln \frac{\Lambda_{\rm UV}}{\mu}\,,
\eeq
where $b_{Y1} = 40/9$ and $\mu < \Lambda_{\rm UV}$. By definition, the Landau pole is reached at scale $\Lambda_L$, where  $1/\alpha_{Y1}(\Lambda_L) = 0$.  If we denote the scale of the symmetry breaking as $\Lambda \sim M_{G'} \sim M_{Z'} < \Lambda_L$, then 
\beq
\frac{1}{\alpha_{Y1}(\Lambda)} = \frac{b_{Y1}}{2\pi} \ln \frac{\Lambda_L}{\Lambda}\, .
\label{const-Landau-pole}
\eeq
The low energy four-fermion coupling $\kappa_1$ is related to the gauge couplings of the $U(1)_{Y_1}$ and $U(1)_{Y}$ groups as $\kappa_1 = \alpha_{Y1} - \alpha_Y $. Then Eq. (\ref{const-Landau-pole}) allows one to determine $\kappa_1$ for given $\Lambda_L/\Lambda$. 

\begin{figure}[htbp]
\begin{center}
\includegraphics{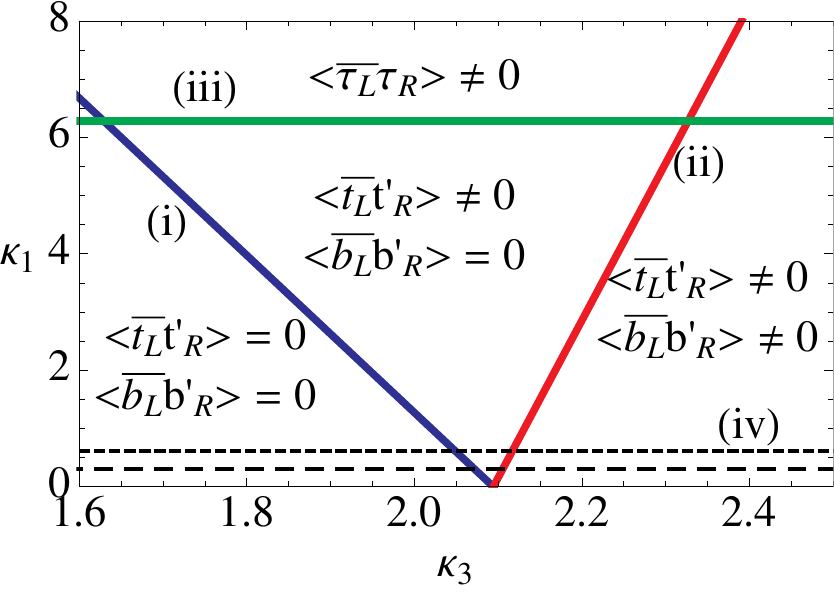} 
\caption{ 
The gap triangle for model B.
The region above (i) represents $\vev{\bar{t}_L t'_R} \neq 0$, the region below (ii) represents $\vev{\bar{b}_L b'_R} \neq 0$ and the region above (iii) represents $\VEV{\tau} \neq 0$. $t_L-t'_R$ and $b_L-b'_R$ form their condensates in an area which is right hand side of (ii) and below (iii). The dahsed lines (iv) represent constraints from a position of the Landau pole, Eq. (\ref{const-Landau-pole}), for $\Lambda_L/\Lambda=10$ (the upper line) and $\Lambda_L/\Lambda=100$ (the lower line). All regions in this figure are allowed by the constraint from $\Delta \rho_*$, Eq.(\ref{const-delta-final}).
\label{gaptriangle-model2}}
\end{center}
\end{figure}

All of these constraints on the parameter space $(\kappa_3, \kappa_1)$ are shown in Fig.\ref{gaptriangle-model2}. The criticality conditions in Eq. (\ref{crit-cond-model2}) result in the gap-triangle and the dashed horizontal line is determined by the constraint on the Landau pole, Eq. (\ref{const-Landau-pole}). The viable region in this plane is at large enough $\kappa_3$ so that the desired condensations are triggered, but at small $\kappa_1$ so that the Landau pole remains sufficiently far in the ultraviolet. The condition (\ref{const-delta-final}) on $\Delta \rho_*$ is weaker than the other constraints as it does not provide additional cuts on the parameter space.

We have assumed that $M_{G^\prime}\sim M_{Z^\prime}$ and that this common scale provides the cutoff $\Lambda$ below which our NJL analysis is valid. Generically one might expect that $\Lambda\sim 1$ TeV, but let us nevertheless discuss in some detail the existing direct observation limits for the coloron and $Z^\prime$-boson. Let us begin with the $Z^\prime$ boson. The contact interaction Eq.(\ref{4fermi-model2-Zprime}) by $Z'$ exchange is limited by the LEP experiments.  In the case of Model B,  if the contact interaction by $Z'$ exchange induces $e^+ e^- \to f \bar{f}$,  the $Z'$ boson mass is given by \cite{Chivukula:2002ry}
\beq
M_{Z'}
=
\Lambda^{{\rm sgn}[Y^{e_i} Y^{f_i}]} 
\sqrt{\frac{\alpha_{\rm em}}{c^2_W} |Y^{e_i} Y^{f_i}|}\,,
\eeq
where $i=L,R$ and $Y^{f_i}$ is the ordinary hypercharge of the fermion $f$. The LEP data provides limits on $\Lambda^\pm$ for various $f$. In the case of Model B, the strongest lower bound comes from $e^+_R e^-_R \to \tau^+_R \tau^-_R $ and the LEP limits on $\Lambda^+_{RR}$ is $\Lambda^+_{RR} > 8.2 \TeV$ at $95\%\cl$ \cite{Alcaraz:2006mx}. This LEP limit implies $M_{Z'} > 798 \GeV$ which is independent of the mixing angle $\theta'$.

Then, let us turn to the case coloron $G'_\mu$ which may be produced at Tevatron 
in the process $p\bar{p} \to G'_\mu \to t\bar{t},b\bar{b},t'\bar{t'},b'\bar{b'}$.
If $M_{G^\prime} < 2m_t,m_t+m_{t'},m_t+m_{b'},2m_{t'},2m_{b'}$, 
the dominant process among them is $p\bar{p} \to G'_\mu \to b\bar{b}$. 
However, only the case of a special coloron (so-called topgluon), which couples strongly to third-family i.e. only the top and bottom quark transform under $SU(3)_1$, has been limited by CDF. The CDF collaboration reported that topgluon mass approximately $300 - 600 \GeV$ is excluded at $95 \% \cl$ for topgluon widths in the $0.3 < \Gamma_C/M_{G'} < 0.7$ \cite{Abe:1998uz}. However, note that the coloron in Model B is different from topgluon and  there are additional vector-like quarks in Model B. Therefore, a more thorough study of various contributions to experimental observables should be carried out in our model; we leave this study for a future work. The case with only the vector-like top quark has been studied in \cite{Dobrescu:2009vz}, however in our case the additional effects coming from vector-like bottom quark need to be included. 

\section{Discussion and conclusions}

In this paper we have constructed explicit models extending the MWT model with topcolor-like dynamics. We considered first the case where a fourth generation of QCD quarks exist and then, second, the case where a fourth generation leptons exists. Both of these possibilities arise due to the requirement that the resulting model is free of global and gauge anomalies. Earlier, MWT has been shown to be viable model for dynamical electroweak symmetry breaking. Here we have extended MWT to address some issues in flavor physics, most importantly the splitting of the top and bottom quark masses by dynamical mechanism, and studied the phenomenological viability of these extensions. 

The first model, termed model A, we considered, illustrates how the mechanisms in topcolor assisted technicolor \cite{Hill:1994hp} allow for a natural explanation of both the top-bottom mass splitting as well as the mass splitting between the bottom quark and the fourth generation quarks so that $t^\prime$ and $b^\prime$ remain heavy to avoid having been observed.

The second model we considered, and called model B, features only fourth family leptons, and top-bottom mass splitting can be explained by the top-seesaw mechanism \cite{Chivukula:1998wd}.
In this model we studied the constraints arising from the requirement of existence of the vacuum condensates required for electroweak symmetry breaking and family of third family quark masses, from the requirement of weak enough $U(1)$ couplings and from the requirement of having only small corrections to $\Delta\rho_\ast$.  

We found both of these possibilities phenomenologically viable in light of the constraints we considered. Several further phenomenological properties are now open for further study both in model A and in model B. In both models the existence of Standard Model -like matter fields provides an immediate handle into the phenomenology. In model A, the fourth generation quarks are interesting for their implications on CP-violation and the CKM paradigm on quark mixing as well as on their direct searches in the LHC experiments. In model B, a similar role is played by the existence of fourth family leptons. The phenomenological viability of new QCD quarks or leptons in light of the electroweak precision data analysis has been studied thoroughly for different mass patterns in \cite{Antipin:2009ks,Antipin:2010it} where also some collider signatures have been considered. Addressing the origin of the mass patterns of the fourth family leptons in model B via a dynamical mechanism analogous to one we have considered here remains a challenge for future work as well as the collider signatures of the vector-like QCD fermions. 

Both in model A and B the contributions to processes mediated by flavor-changing neutral current interactions are expected to be small due to the flavor universality of the color sector. In general the phenomenology of Model B is expected to resemble the phenomenology of the top-quark seesaw model \cite{He:2001fz}. Model B also features possible interesting dark matter candidates. In particular, the fourth family neutrino, if stable, is a natural candidate WIMP and within the standard annihilation versus thermal freezeout scenario was the neutrino was shown to be a viable candidate for cold dark matter provided some Dirac-Majorana type mass mixing is present \cite{Kainulainen:2009rb}. 


\end{document}